\documentclass{elsartL}
\usepackage{graphicx}
\usepackage{amssymb}
\usepackage{amsmath}
\usepackage{appendix}
\usepackage{mathrsfs}
\usepackage{mathtools}
\newcommand{\fsl}[1]{\ensuremath{\mathrlap{\!\not{\phantom{#1}}}#1}}
\begin{document}

\begin{frontmatter}
\title{The propagation of a massive spin-$\frac{3}{2}$ field, with application to $\pi N$ scattering}
\author[EM]{E. Matsinos{$^*$}},
\author[GR]{G. Rasche},
\address[EM]{Institute of Mechatronic Systems, Zurich University of Applied Sciences, Technikumstrasse 5, CH-8401 Winterthur, Switzerland}
\address[GR]{Physik-Institut der Universit\"at Z\"urich, Winterthurerstrasse 190, CH-8057 Z\"urich, Switzerland}

\begin{abstract}
In the present paper, we investigate the propagation of a massive spin-$\frac{3}{2}$ field, aiming at a direct application in hadronic models of the pion-nucleon ($\pi N$) interaction. Suitable expressions for the contributions to 
the standard invariant amplitudes $A$ and $B$ are derived, applicable in the general case of isospin decomposition. We first deal with the details of the lengthy calculation involving the Rarita-Schwinger propagator and confirm the 
validity of the expressions which had appeared in the literature in the early 1970s. We subsequently derive the corresponding contributions when following two other approaches, one featuring the Williams propagator, the other being 
known as Pascalutsa's method.
%
%
\end{abstract}
{$^*$}{Corresponding author. E-mail: evangelos[DOT]matsinos[AT]zhaw[DOT]ch, evangelos[DOT]matsinos[AT]sunrise[DOT]ch}
\end{frontmatter}

\section{\label{sec:Intro}Introduction}

The treatment of Feynman graphs (simply `graphs' hereafter), involving virtual fermions with spin $J>\frac{1}{2}$, is rather intricate; as a result of the off-shellness of the intermediate state, its propagator is expected to 
contain, apart from the `nominal' contributions of the spin-$J$ state, contributions from the states of lower spin $J-n$, where $1 \leq n \leq J-\frac{1}{2}$. As the fixation of their admixture can hardly be made on a theoretical 
basis, each of these states introduces into the problem one additional free parameter. Furthermore, the effort, needed for obtaining the contributions to the scattering amplitude, increases rapidly with $J$.

The aim of the present paper is to revisit the subject of the contributions (to the scattering amplitude) of graphs involving a massive spin-$\frac{3}{2}$ intermediate state and provide details which had been omitted in relevant 
past papers. Although our focus (and main interest) lies with an application to pion-nucleon ($\pi N$) scattering, the generalisation of the approach to two-hadron scattering may be achieved by changing the masses of the 
interacting particles and including the appropriate isospin decomposition of the scattering amplitude. Regarding the $\pi N$ system, below a typical (pion laboratory kinetic) energy of a few hundred MeV, the contributions to the 
hadronic part of the scattering amplitude of the graphs with a $\Delta(1232)$ intermediate state (Figs.~\ref{fig:PiNDelta}) are indispensable in the description of the experimental data.

In the present paper, we will make use of the following notation and conventions.
\begin{itemize}
\item The speed of light in vacuum $c$ is equal to $1$.
\item Einstein's summation convention is used.
\item $I_n$ denotes the $n \times n$ identity matrix.
\item $g^{\mu \nu}$ denotes the Minkowski metric with signature `$+ \, - \, - \, -$'.
\item The isospin operators of the nucleon and of the pion are denoted by $\frac{1}{2} \vec{\tau}$ and $\vec{t}$.
\item $\gamma^{\mu}$ ($\mu = 0, 1, 2, 3$) are the standard Dirac $4 \times 4$ matrices, satisfying the relation $\{ \gamma^{\mu}, \gamma^{\nu} \} = 2 g^{\mu \nu} I_4$.
\item $m_p$ and $m_c$ denote the masses of the proton and of the charged pion.
\item $s$, $u$, and $t$ are the standard Mandelstam variables; we also make use of the Mandelstam variable $\nu$, defined as
\begin{equation} \label{eq:EQ000}
\nu=\frac{s-u}{4 m_p} \, \, \, .
\end{equation}
\item For a $4$-vector $a$, $\fsl{a}=\gamma^{\mu} a_{\mu}$; the corresponding $3$-vector (i.e., the vector of the spatial components of the $4$-vector $a$) is denoted by $\vec{a}$.
\item $T$ stands for the pion laboratory kinetic energy. Supposing that the nucleon (a proton target is assumed) is initially at rest in the laboratory system, the relation between $s$ and $T$ is: $s = (m_p + m_c)^2 + 2 T m_p$.
\item CM stands for the centre of mass.
\item The total energy $W$ in the CM frame obeys the relation: $s=W^2$.
\item $p$ and $q$ are the $4$-momenta of the nucleon and of the incident pion in the CM frame, which is defined by $\vec{p} + \vec{q} = \vec{0}$.
\item $p^\prime$ and $q^\prime$ are the $4$-momenta of the scattered nucleon and of the scattered pion in the CM frame; of course, $\vec{p} \, ^{\prime} + \vec{q} \, ^{\prime} = \vec{0}$.
\item $\theta$ denotes the scattering angle in the CM frame.
\end{itemize}

Energy-momentum conservation enforces the relation $p + q = p^\prime + q^\prime$. For elastic scattering, $q_0 = q^\prime_0$ (consequently, $p_0 = p^\prime_0$).

The present paper is organised as follows. Section \ref{sec:piND} provides a summary of the historical developments regarding the propagation of $\Delta(1232)$ in the context of the $\pi N$ interaction. Section \ref{sec:Derivation} 
provides useful details for the calculation of the corresponding contributions to the scattering amplitude, the separation of the effects into pole and non-pole parts, and the isospin decomposition of the resulting amplitudes. In 
Sections \ref{sec:RS}, \ref{sec:WL}, and \ref{sec:PL}, details are given regarding the lengthy calculation of the contributions to the standard invariant amplitudes $A$ and $B$, using three approaches, featuring the Rarita-Schwinger 
propagator, the propagator which Williams introduced in the mid 1980s, and the method which Pascalutsa proposed in the late 1990s; suitable expressions are given, applicable in the general case of isospin decomposition of the 
scattering amplitude. In Section \ref{sec:Comparison}, we compare the invariant amplitudes, obtained with these three approaches, and briefly discuss a few obvious differences. Section \ref{sec:Conclusions} contains a summary of 
the main findings of the present work.

\section{\label{sec:piND}The $\Delta(1232)$ and the $\pi N \Delta$ interaction}

The need to determine the contributions to the scattering amplitude of graphs with a $\Delta(1232)$ intermediate state originally served as the main motivation for developing methods of treating the propagation of massive 
spin-$\frac{3}{2}$ fields. It thus appears natural to unfold the relevant historical developments bearing the $\Delta(1232)$ in mind.

The first attempts to construct the massive spin-$\frac{3}{2}$ propagator date back to the late 1930s \cite{fp} and early 1940s \cite{rs}. The propagator, then obtained, has been known in the literature as the `Rarita-Schwinger 
propagator'; it comprises spin-$\frac{3}{2}$ and spin-$\frac{1}{2}$ contributions, and contains an arbitrary (in general, complex) parameter $A \neq -\frac{1}{2}$. In the mid 1980s, Williams proposed a propagator which did not 
contain spin-$\frac{1}{2}$ components \cite{wl}, but shortly afterwards Benmerrouche, Davidson, and Mukhopadhyay \cite{bdm} argued that, as it has no inverse, the `Williams propagator' cannot be correct. Other propagators appeared 
in the late 1990s, without \cite{pasca1,pasca2} and with \cite{haber} spin-$\frac{1}{2}$ contributions. In the analyses after (and including) Ref.~\cite{glm}, we have followed the Rarita-Schwinger formalism, as presented in 
Ref.~\cite{h}, with $A=-1$. (This choice for $A$ eliminates part of the spin-$\frac{1}{2}$ contributions.)

The interaction Lagrangian density
\begin{equation} \label{eq:EQ001}
\Delta \mathscr{L}_{\pi N \Delta} = \frac{g_{\pi N \Delta}}{2 m_p} \bar{\Psi}_\mu \vec{T} \cdot \Theta^{\mu \nu} \partial_\nu \vec{\pi} \psi + h.c.
\end{equation}
introduces two parameters: the coupling constant~\footnote{When comparing values of the coupling constant $g_{\pi N \Delta}$, the reader must bear in mind that, in some works (e.g., in Ref.~\cite{h}), the factor $2 m_p$, appearing 
as denominator on the right-hand side (rhs) of Eq.~(\ref{eq:EQ001}), is absorbed in $g_{\pi N \Delta}$. One possibility of fixing the coupling constant $g_{\pi N \Delta}$ from the width of the $\Delta(1232)$ is discussed in 
Subsection 3.5.1 of Ref.~\cite{mr} (see footnote 10 therein).} $g_{\pi N \Delta}$ and the parameter $Z$ associated with the vertex factor
\begin{equation} \label{eq:EQ002}
\Theta^{\mu \nu} = g^{\mu \nu} - \left( Z + \frac{1}{2} \right) \gamma^\mu \gamma^\nu \equiv g^{\mu \nu} + z \gamma^\mu \gamma^\nu \, \, \, .
\end{equation}
In Eq.~(\ref{eq:EQ001}), $\Psi_\mu$ stands for the spinor-vector field of the $\Delta(1232)$; the spinor index is suppressed. $\vec{T}$ is the transition operator between the total-isospin $I=\frac{3}{2}$ and $I=\frac{1}{2}$ states. 
The parameter $Z$ has also been associated with the spin-$\frac{1}{2}$ contributions to the $\Delta(1232)$ field~\footnote{Some authors \cite{pasca2,te,kem} have argued that the spin-$\frac{1}{2}$ contributions to the $\Delta(1232)$ 
field are redundant in the framework of an Effective Field Theory, as such off-shell effects can be absorbed in other terms of the effective Lagrangian. Nevertheless, this claim does not impair the possibility of including these 
contributions in phenomenological models.}, entering the scattering amplitude via the $\pi N \Delta$ interaction vertex (rather than directly, i.e., via the $\Delta(1232)$ propagator). In general, $Z$ is complex \cite{h}, but herein 
it will be assumed to be real; we are not aware of works, in which the quantities $A$ and $Z$ have been treated as complex parameters.

The fixation of the parameter $Z$ from theoretical principles has been explored in a number of studies. To start with, using the subsidiary condition $\gamma^\mu \Theta_{\mu \nu} = 0$, Peccei suggested the use of $Z = -\frac{1}{4}$ 
\cite{pe}, whereas Nath, Etemadi, and Kimel \cite{nek} recommended $Z=\frac{1}{2}$. However, it was argued in Ref.~\cite{bdm} that the $Z=\frac{1}{2}$ choice leads to unexpected properties of the $\Delta$ radiative decay. H\"ohler 
\cite{h} assumed a cautious attitude regarding the arguments in favour of such `theoretical preferences', thus hinting at the extraction of the value of $Z$ from measurements. Although $Z$ has been treated as a free parameter thus 
far, the results of our partial-wave analyses (PWAs) of the $\pi^\pm p$ elastic-scattering data favour the case $Z = -\frac{1}{2}$; this value leads to the simplification of the relevant expressions and considerably shortens the 
calculations within the Rarita-Schwinger formalism. This observation has two consequences: a) It leaves open the possibility of fixing $Z$ to $-\frac{1}{2}$ (hence $z$ to $0$) in the future, thus of performing the fits (to the 
$\pi N$ scattering data) with fewer parameters. b) It may be of relevance in terms of the treatment of graphs involving states of higher spin, i.e., of the treatment of the massive spin-$\frac{5}{2}$ and spin-$\frac{7}{2}$ fields.

\section{\label{sec:Derivation}Derivation of the contributions to the invariant amplitudes $A$ and $B$}

Leaving out, for a moment, the isospin decomposition of the $\pi N$ scattering amplitude, the contributions of a graph (of a hadronic model) to the $T$-matrix element in the CM frame are put in the form:
\begin{equation} \label{eq:EQ003}
\mathscr{T} = \bar{u}_f (p^\prime) \, \left( A + B (\gamma^{0} W - m_p) \right) \, u_i (p) \, \, \, ,
\end{equation}
where $u (p)$ is the Dirac spinor associated with the plane-wave of a nucleon with $4$-momentum $p$; $\bar{u} (p) = u^\dagger (p) \gamma^0$ is the conjugate spinor, whereas $u^\dagger (p)$ is the conjugate transpose of $u (p)$. 
The subscripts $i$ and $f$ refer to the nucleon spin and isospin in the initial and final state, respectively. The functions $A$ and $B$ in Eq.~(\ref{eq:EQ003}) represent the contributions of the specific graph to the scattering 
amplitude; $A$ and $B$ are functions of two (independent) Mandelstam variables, which may be chosen at will. To facilitate the comparison of our expressions with other works, chosen herein are the Mandelstam variables $s$ and $t$. 
A thorough description of the formalism, used in the present paper, may be found in Ref.~\cite{mr}; the isospin-even or isoscalar (denoted by the superscript `$+$') and isospin-odd or isovector (denoted by the superscript `$-$') 
amplitudes are defined therein, in the beginning of Section 3.

The $s$-channel contribution to the $T$-matrix element in the CM frame reads as:
\begin{equation} \label{eq:EQ004}
\mathscr{T} = -i \left( \frac{g_{\pi N R}}{2 m_p} \right) ^2 \bar{u}_f (p^\prime) ( q^{\prime \, \mu} + z \fsl{q}^{\, \prime} \gamma^\mu) \,\, \Pi_{\mu \nu} (P) \,\, (q^\nu + z \gamma^\nu \fsl{q} ) u_i (p) \, \, \, ,
\end{equation}
where the propagator $\Pi_{\mu \nu} (P)$ of the massive spin-$\frac{3}{2}$ intermediate state (denoted as $R$) has the form:
\begin{equation} \label{eq:EQ005}
\Pi_{\mu \nu} (P) = \frac{i}{\fsl{P} - m_R} \mathscr{P}_{\mu \nu} (P) \, \, \, ;
\end{equation}
in Eqs.~(\ref{eq:EQ004}) and (\ref{eq:EQ005}), $P$ denotes the $4$-momentum of the massive spin-$\frac{3}{2}$ intermediate state (in the $s$ channel, $P=p+q$), $m_R$ its mass, and $\mathscr{P}_{\mu \nu} (P)$ the spin-projection 
operator associated with the propagator $\Pi_{\mu \nu} (P)$. Using Eq.~(\ref{eq:EQ005}), along with the definition of the Mandelstam variable $s$, one may rewrite Eq.~(\ref{eq:EQ004}) as:
\begin{equation} \label{eq:EQ006}
\mathscr{T} = \frac{1}{s - m_R^2} \left( \frac{g_{\pi N R}}{2 m_p} \right) ^2 \bar{u}_f (p^\prime) ( q^{\prime \, \mu} + z \fsl{q}^{\, \prime} \gamma^\mu) (\fsl{p}+\fsl{q}+m_R) \,\, \mathscr{P}_{\mu \nu} (p+q) \,\, (q^\nu + z \gamma^\nu \fsl{q} ) u_i (p) \, \, \, .
\end{equation}

The $u$-channel contributions to the $T$-matrix element in the CM frame may be derived from Eq.~(\ref{eq:EQ006}) via the standard substitutions: $s \rightarrow u$, $q \rightarrow -q^\prime$, and $q^\prime \rightarrow -q$.
\begin{equation} \label{eq:EQ007}
\mathscr{T} = \frac{1}{u - m_R^2} \left( \frac{g_{\pi N R}}{2 m_p} \right) ^2 \bar{u}_f (p^\prime) (q^\mu + z \fsl{q} \gamma^\mu) (\fsl{p}-\fsl{q}^{\, \prime}+m_R) \,\, \mathscr{P}_{\mu \nu} (p-q^\prime) \,\, (q^{\prime \, \nu} + z \gamma^\nu \fsl{q}^{\, \prime} ) u_i (p)
\end{equation}

Regarding the application to the $\pi N$ system, the intermediate state may have total isospin $I=\frac{3}{2}$ or $I=\frac{1}{2}$. Therefore, the massive spin-$\frac{3}{2}$ intermediate state may be either an 
$I(J^P) = \frac{3}{2}(\frac{3}{2}^+)$ state (e.g., the $\Delta(1232)$) or an $I(J^P) = \frac{1}{2}(\frac{3}{2}^+)$ state (e.g., the $N(1720)$). In the former case, the isospin decomposition of the scattering amplitude involves 
the combination $\frac{2}{3} + \frac{1}{3} \vec{\tau} \cdot \vec{t}$ for the $s$-channel graph and $\frac{2}{3} - \frac{1}{3} \vec{\tau} \cdot \vec{t}$ for the $u$-channel graph; in the latter case, the combinations are: 
$1 - \vec{\tau} \cdot \vec{t}$ for the $s$-channel graph and $1 + \vec{\tau} \cdot \vec{t}$ for the $u$-channel graph. To facilitate the use of our results, expressions will be derived for the general isospin decomposition, 
$\alpha + \beta \vec{\tau} \cdot \vec{t}$ for the $s$-channel graphs and $\alpha - \beta \vec{\tau} \cdot \vec{t}$ for the $u$-channel graphs. To adapt our expressions to $\pi N$ scattering, one must use $\alpha=\frac{2}{3}$ and 
$\beta=\frac{1}{3}$ for the contributions of the $\Delta(1232)$ graphs (Figs.~\ref{fig:PiNDelta}), and $\alpha=1$ and $\beta=-1$ for the (significantly smaller) contributions of the $N(1720)$ graphs.

The separation of the contributions to the scattering amplitude into pole and non-pole parts is rather subtle. In the framework of Refs.~\cite{h,nek}, the $s$-channel contributions to the invariant amplitudes $A$ and $B$ are 
functions of the Mandelstam variables $s$ and $t$, and may be put in the form of a sum of three terms, each containing a different power of $s-m_R^2$, from $-1$ (inversely proportional) to $1$ (linear); the pole contributions 
comprise only the terms containing $(s-m_R^2)^{-1}$. (Regarding the $u$-channel contributions, the previous comment holds after substituting $s$ with $u$.) The separation of the contributions into pole and non-pole parts depends 
on the choice of the independent variables in the problem. For instance, if one chooses to use $s$ and $\cos\theta$ (instead of $s$ and $t$), some terms categorised within the pole part will be transferred to the non-pole 
contributions. As earlier mentioned, the choice of $s$ and $t$ as independent variables in the present paper facilitates the comparison of our results with the standard literature, namely with the expressions of Refs.~\cite{h,nek} 
(for the Rarita-Schwinger propagator).

\section{\label{sec:RS}The Rarita-Schwinger propagator}

In the Rarita-Schwinger formalism, the operator $\mathscr{P}_{\mu \nu} (P)$ has the form:
\begin{align} \label{eq:EQ008}
\mathscr{P}_{\mu \nu}^{RS} (P) = g_{\mu \nu} - \frac{1}{3} \gamma_\mu \gamma_\nu - \frac{\gamma_\mu P_\nu - \gamma_\nu P_\mu}{3 m_R} - \frac{2 P_\mu P_\nu}{3 m_R^2} \, \, \, .
\end{align}

Before advancing to the technicalities of the calculation, we will briefly comment on the structure of the Rarita-Schwinger propagator. To this end, we will follow the established approach and put $\mathscr{P}_{\mu \nu}^{RS} (P)$ 
in the form:
\begin{align} \label{eq:EQ008_1}
\mathscr{P}_{\mu \nu}^{RS} (P) = & \mathscr{P}_{\mu \nu}^{3/2} (P) - \frac{2 (P^2-m_R^2)}{3 m_R^2} (\mathscr{P}^{1/2}_{22})_{\mu \nu} (P) \nonumber \\
& + \frac{\fsl{P}-m_R}{\sqrt{3} m_R} \left( (\mathscr{P}^{1/2}_{12})_{\mu \nu} (P) + (\mathscr{P}^{1/2}_{21})_{\mu \nu} (P) \right) \, \, \, ,
\end{align}
where all the expressions for the operators $(\mathscr{P}^{1/2}_{mn})_{\mu \nu} (P)$, as well as for $\mathscr{P}_{\mu \nu}^{3/2} (P)$, may be found in Ref.~\cite{haber}. $\mathscr{P}_{\mu \nu}^{3/2} (P)$ projects onto pure 
spin-$\frac{3}{2}$ states and is denoted in Ref.~\cite{haber} as $\mathcal{D}_{\mu \nu} (P)$ (see Eq.~(7) therein), whereas $(\mathscr{P}^{1/2}_{12})_{\mu \nu} (P)$ and $(\mathscr{P}^{1/2}_{21})_{\mu \nu} (P)$ describe the 
transitions between the two irreducible spin-$\frac{1}{2}$ representations (see Eqs.~(10a) and (10b) therein); finally, the operator $(\mathscr{P}^{1/2}_{22})_{\mu \nu} (P)$ (defined in Eq.~(9) of Ref.~\cite{haber}) is the 
projector associated with the ($0$,$\frac{1}{2}$) irreducible representation. It follows that $\mathscr{P}_{\mu \nu}^{RS} (P) \equiv \mathscr{P}_{\mu \nu}^{3/2} (P)$ on the mass shell ($P^2=m_R^2$).

In Subsections \ref{sec:RSs} and \ref{sec:RSu}, explicit results will be given for the contribution of each of the four terms, appearing on the rhs of Eq.~(\ref{eq:EQ008}), to the invariant amplitudes $A$ and $B$. The calculations 
are somewhat lengthy, but straightforward, assuming familiarity with the algebra of the Dirac matrices. The relations of Appendix \ref{App:AppA} are helpful in the calculation. The contributions to the invariant amplitudes $A$ and 
$B$ can be disentangled easily, as the latter involve the factor $(\gamma^{0} W - m_p)$ (see Eq.~(\ref{eq:EQ003})).

Prior to entering the details of the calculation, it must be mentioned that the expressions for the pole and non-pole contributions of the $\Delta(1232)$ graphs (Figs.~\ref{fig:PiNDelta}) to the invariant amplitudes $A$ and $B$ 
may be found in Ref.~\cite{h}, pp.~562 and 564. Due to a sign-convention difference, the contributions to the isovector invariant amplitudes $A^-$ and $B^-$ of the present paper are opposite to those of Ref.~\cite{h}. The isoscalar 
invariant amplitudes $A^+$ and $B^+$ have the same sign. The expressions had appeared earlier in Ref.~\cite{nek}, but the concise formulae of Ref.~\cite{h} are more attractive for a compact implementation~\footnote{The reader must 
also bear in mind that, compared to Ref.~\cite{h}, the invariant amplitudes $B$ are defined with an opposite sign in Ref.~\cite{nek}.}. The main goal in the present section is the verification of the expressions given in 
Refs.~\cite{h,nek}.

\subsection{\label{sec:RSs}The $s$-channel contributions}

As mentioned at the end of Section \ref{sec:Derivation}, the contributions will be split into pole (inversely proportional to $s - m_R^2$) and non-pole (all else) terms. For the sake of brevity, in the contributions of Subsections 
\ref{sec:RSs1}-\ref{sec:RSs4}, the factor $\frac{1}{s - m_R^2} \left( \frac{g_{\pi N R}}{2 m_p} \right) ^2$, appearing on the rhs of Eq.~(\ref{eq:EQ006}), is suppressed.

\subsubsection{\label{sec:RSs1}The contributions of the first term}

Pole contributions to the invariant amplitude $A$
\begin{equation} \label{eq:EQ009}
a_{s,p;1} = (m_R + m_p) (m_c^2 - \frac{t}{2}) + 2 z \big[ (1 + 2 z) m_R + (1 + z) m_p \big] (m_R^2 - m_p^2)
\end{equation}
Pole contributions to the invariant amplitude $B$
\begin{equation} \label{eq:EQ010}
b_{s,p;1} = m_c^2 - \frac{t}{2} - 2 z (2 m_R m_p + 2 m_p^2 - m_c^2) - 2 z^2 (m_R^2 + 4 m_R m_p + m_p^2)
\end{equation}
Non-pole contributions to the invariant amplitude $A$
\begin{equation} \label{eq:EQ011}
a_{s,np;1} = 2 z \big[ (1 + 2 z) m_R + (1 + z) m_p \big] (s - m_R^2)
\end{equation}
Non-pole contributions to the invariant amplitude $B$
\begin{equation} \label{eq:EQ012}
b_{s,np;1} = - 2 z^2 (s - m_R^2)
\end{equation}

\subsubsection{\label{sec:RSs2}The contributions of the second term}

Pole contributions to the invariant amplitude $A$
\begin{equation} \label{eq:EQ013}
a_{s,p;2} = - \frac{1 + 4z}{3} \big[ (1 + 4 z) m_R + (1 + 2 z) m_p \big] (m_R^2 - m_p^2)
\end{equation}
Pole contributions to the invariant amplitude $B$
\begin{equation} \label{eq:EQ014}
b_{s,p;2} = \frac{1 + 4z}{3} \big[ 2 m_R m_p + 2 m_p^2 - m_c^2 + 2 z (m_R^2 + 4 m_R m_p + m_p^2) \big]
\end{equation}
Non-pole contributions to the invariant amplitude $A$
\begin{equation} \label{eq:EQ015}
a_{s,np;2} = - \frac{1 + 4z}{3} \big[ (1 + 4 z) m_R + (1 + 2 z) m_p \big] (s - m_R^2)
\end{equation}
Non-pole contributions to the invariant amplitude $B$
\begin{equation} \label{eq:EQ016}
b_{s,np;2} = \frac{1 + 4z}{3} 2 z (s - m_R^2)
\end{equation}

\subsubsection{\label{sec:RSs3}The contributions of the third term}

Pole contributions to the invariant amplitude $A$
\begin{equation} \label{eq:EQ017}
a_{s,p;3} = \frac{1}{3 m_R} \Big\{ - \frac{m_c^4}{2} + \big[ (\frac{1}{2} + 4 z + 6 z ^2) m_R^2 - (\frac{1}{2} + 2 z) m_p^2 + 2 z m_c^2 \big] (m_R^2 - m_p^2) \Big\}
\end{equation}
Pole contributions to the invariant amplitude $B$
\begin{equation} \label{eq:EQ018}
b_{s,p;3} = - \frac{m_p}{3 m_R} \big[ (1 + 6z + 12z^2) m_R^2 - (1+2z) (m_p^2 - m_c^2) \big]
\end{equation}
Non-pole contributions to the invariant amplitude $A$
\begin{align} \label{eq:EQ019}
a_{s,np;3} = \frac{s - m_R^2}{3 m_R} \big[ & (1 + 8 z + 12 z^2) m_R^2 - (1 + 6 z + 6 z^2) m_p^2 + 2 z m_c^2 \nonumber \\
&+ (\frac{1}{2} + 4 z + 6 z^2) (s - m_R^2) \big]
\end{align}
Non-pole contributions to the invariant amplitude $B$
\begin{equation} \label{eq:EQ020}
b_{s,np;3} = - \frac{s - m_R^2}{3 m_R} (1 + 6 z + 12 z^2) m_p
\end{equation}

\subsubsection{\label{sec:RSs4}The contributions of the fourth term}

Pole contributions to the invariant amplitude $A$
\begin{align} \label{eq:EQ021}
a_{s,p;4} = - \frac{2 (m_R + m_p)}{3 m_R^2} \big[ & \frac{(m_R^2 - m_p^2 + m_c^2)^2}{4} + z m_R (m_R - m_p) (m_R^2 - m_p^2 + m_c^2) \nonumber \\
&+ z^2 m_R^2 (m_R - m_p)^2 \big]
\end{align}
Pole contributions to the invariant amplitude $B$
\begin{align} \label{eq:EQ022}
b_{s,p;4} = - \frac{2}{3 m_R^2} \big[ & \frac{(m_R^2 - m_p^2 + m_c^2)^2}{4} + z m_R (m_R - m_p) (m_R^2 -m_p^2 + m_c^2) \nonumber \\
&+ z^2 m_R^2 (m_R - m_p)^2 \big]
\end{align}
Non-pole contributions to the invariant amplitude $A$
\begin{align} \label{eq:EQ023}
a_{s,np;4} = - \frac{2 (s - m_R^2)}{3 m_R^2} \Big\{ & \frac{m_R + m_p}{2} (m_R^2 - m_p^2 + m_c^2) + z m_R (2 m_R^2 - 2 m_p^2 + m_c^2) \nonumber \\
&+ z^2 (m_R - m_p) (2 m_R^2 - m_p^2) \nonumber \\
&+ \big[ (\frac{1}{4} + z + z^2) m_R + (\frac{1}{4} - z^2) m_p \big] (s - m_R^2) \Big\}
\end{align}
Non-pole contributions to the invariant amplitude $B$
\begin{align} \label{eq:EQ024}
b_{s,np;4} = - \frac{2 (s - m_R^2)}{3 m_R^2} \big[ & \frac{m_R^2 - m_p^2 + m_c^2}{2} + z (2 m_R^2 - m_R m_p - m_p^2 + m_c^2) \nonumber \\
&+ z^2 (2 m_R^2 - 2 m_R m_p + m_p^2) \nonumber \\
&+ (\frac{1}{4} + z + z^2) (s - m_R^2) \big]
\end{align}

\subsubsection{\label{sec:RSsTotal}Sums of the $s$-channel contributions}

The $s$-channel contributions to the invariant amplitudes $A$ and $B$ may be obtained from the results listed in Subsections \ref{sec:RSs1}-\ref{sec:RSs4} via the expressions.
\begin{equation} \label{eq:EQ025}
A_{s,p} = \frac{1}{s - m_R^2} \left( \frac{g_{\pi N R}}{2 m_p} \right) ^2 \sum_{n=1}^4 a_{s,p;n}
\end{equation}
\begin{equation} \label{eq:EQ026}
B_{s,p} = \frac{1}{s - m_R^2} \left( \frac{g_{\pi N R}}{2 m_p} \right) ^2 \sum_{n=1}^4 b_{s,p;n}
\end{equation}
\begin{equation} \label{eq:EQ027}
A_{s,np} = \frac{1}{s - m_R^2} \left( \frac{g_{\pi N R}}{2 m_p} \right) ^2 \sum_{n=1}^4 a_{s,np;n}
\end{equation}
\begin{equation} \label{eq:EQ028}
B_{s,np} = \frac{1}{s - m_R^2} \left( \frac{g_{\pi N R}}{2 m_p} \right) ^2 \sum_{n=1}^4 b_{s,np;n}
\end{equation}

After some algebraic operations, one obtains:
\begin{align} \label{eq:EQ029}
\sum_{n=1}^4 a_{s,p;n} = - \frac{1}{3} \big[ & \frac{3 (m_R + m_p) t}{2} + 4 m_R p_{0 R}^2 + 2 m_R m_p p_{0 R} - 2 m_R m_p^2 \nonumber \\
&- 4 m_p^3 - 2 m_p^2 p_{0 R} + 2 m_p p_{0 R}^2 \big] \, \, \, ,
\end{align}
where $p_{0 R}=\frac{m_R^2 + m_p^2 -m_c^2}{2 m_R}$ denotes the total CM energy of the nucleon at the resonance position ($s = m_R^2$). Interestingly, the $Z$-dependence disappears when summing up the pole contributions to the 
invariant amplitude $A$, detailed in Eqs.~(\ref{eq:EQ009}), (\ref{eq:EQ013}), (\ref{eq:EQ017}), and (\ref{eq:EQ021}). In fact, the $Z$-independence is a common characteristic of \emph{all} pole contributions of Section \ref{sec:RS}; 
in this respect, all pole contributions to the invariant amplitudes $A$ and $B$ are unique.

Expression (\ref{eq:EQ029}) can be simplified further. The following compact formula for $A_{s,p}$ may be obtained after including the isospin structure \cite{h}:
\begin{equation} \label{eq:EQ030}
A_{s,p}^\pm = \binom{\alpha}{\beta} \frac{g_{\pi N R}^2}{12 m_p^2} \frac{\alpha_1 + \alpha_2 t}{m_R^2 - s} \, \, \, ,
\end{equation}
with
\begin{align} \label{eq:EQ031}
\alpha_1 &= 3 (m_R + m_p) \vec{q}_R \, ^2 + (m_R - m_p) (p_{0R} + m_p)^2 \, \, \, , \nonumber \\
\alpha_2 &= \frac{3}{2} (m_R + m_p) \, \, \, ,
\end{align}
where $\lvert \vec{q}_R \rvert$ is the (modulus of the) CM momentum at $s=m_R^2$. In particular, for the $s$-channel graph with a $\Delta(1232)$ intermediate state, one obtains:
\begin{equation} \label{eq:EQ032}
A_{s,p}^\pm = \binom{2}{1} \frac{g_{\pi N \Delta}^2}{36 m_p^2} \frac{\alpha_1 + \alpha_2 t}{m_\Delta^2 - s} \, \, \, .
\end{equation}

The sum on the rhs of Eq.~(\ref{eq:EQ026}) reads as:
\begin{equation} \label{eq:EQ033}
\sum_{n=1}^4 b_{s,p;n} = - \frac{1}{3} \big( \frac{3 t}{2} - 4 m_p^2 - 2 m_p p_{0R} + 2 p_{0R}^2 \big) \, \, \, ,
\end{equation}
which leads to
\begin{equation} \label{eq:EQ034}
B_{s,p}^\pm = \binom{\alpha}{\beta} \frac{g_{\pi N R}^2}{12 m_p^2} \frac{\beta_1 + \beta_2 t}{m_R^2 - s} \, \, \, ,
\end{equation}
with
\begin{align} \label{eq:EQ035}
\beta_1 &= 3 \vec{q}_R \, ^2 - (p_{0R} + m_p)^2 \, \, \, , \nonumber \\
\beta_2 &= \frac{3}{2} \, \, \, .
\end{align}
In particular, for the $s$-channel graph with a $\Delta(1232)$ intermediate state, one obtains:
\begin{equation} \label{eq:EQ036}
B_{s,p}^\pm = \binom{2}{1} \frac{g_{\pi N \Delta}^2}{36 m_p^2} \frac{\beta_1 + \beta_2 t}{m_\Delta^2 - s} \, \, \, .
\end{equation}

The final expression for $A_{s,np}$ reads as:
\begin{align} \label{eq:EQ037}
A_{s,np}^\pm = \binom{\alpha}{\beta} \frac{g_{\pi N R}^2}{12 m_p^2} \Big\{ &2 (-1 + z + 2 z^2) m_R + (-3 + 2 z^2) m_p - (1 + 2 z + 4 z^2) \frac{m_p^2}{m_R} \nonumber \\
&+ 2 p_{0R} + 2 \frac{m_p}{m_R} p_{0R} - 2 z^2 \frac{m_p^3}{m_R^2} \nonumber \\
&+ \frac{s - m_R^2}{m_R} \big[ 2 z + 4 z^2 - (\frac{1}{2} - 2 z^2) \frac{m_p}{m_R} \big] \Big\} \, \, \, .
\end{align}

The final expression for $B_{s,np}$ reads as:
\begin{align} \label{eq:EQ038}
B_{s,np}^\pm = - \binom{\alpha}{\beta} \frac{g_{\pi N R}^2}{12 m_p^2} \big[ & 1 + 2z + 2z^2 + (1 + 4z + 8z^2) \frac{m_p}{m_R} \nonumber \\
&- (1 + 2z - 2z^2) \frac{m_p^2}{m_R^2} - 2 \frac{m_c^2}{m_R^2} Z + 2 \frac{s - m_R^2}{m_R^2} Z^2 \big] \, \, \, .
\end{align}

\subsection{\label{sec:RSu}The $u$-channel contributions}

Similarly to Subsection \ref{sec:RSs}, the $u$-channel contributions will be split into pole (inversely proportional to $u - m_R^2$) and non-pole (all else) terms. In the contributions of Subsections \ref{sec:RSu1}-\ref{sec:RSu4}, 
the factor $\frac{1}{u - m_R^2} \left( \frac{g_{\pi N R}}{2 m_p} \right) ^2$, appearing on the rhs of Eq.~(\ref{eq:EQ007}), is suppressed.

\subsubsection{\label{sec:RSu1}The contributions of the first term}

Pole contributions to the invariant amplitude $A$
\begin{equation} \label{eq:EQ039}
a_{u,p;1} = a_{s,p;1}
\end{equation}
Pole contributions to the invariant amplitude $B$
\begin{equation} \label{eq:EQ040}
b_{u,p;1} = - b_{s,p;1}
\end{equation}
Non-pole contributions to the invariant amplitude $A$
\begin{equation} \label{eq:EQ041}
a_{u,np;1} = 2 z \big[ (1 + 2 z) m_R + (1 + z) m_p \big] (u - m_R^2)
\end{equation}
Non-pole contributions to the invariant amplitude $B$
\begin{equation} \label{eq:EQ042}
b_{u,np;1} = 2 z^2 (u - m_R^2)
\end{equation}

\subsubsection{\label{sec:RSu2}The contributions of the second term}

Pole contributions to the invariant amplitude $A$
\begin{equation} \label{eq:EQ043}
a_{u,p;2} = a_{s,p;2}
\end{equation}
Pole contributions to the invariant amplitude $B$
\begin{equation} \label{eq:EQ044}
b_{u,p;2} = - b_{s,p;2}
\end{equation}
Non-pole contributions to the invariant amplitude $A$
\begin{equation} \label{eq:EQ045}
a_{u,np;2} = - \frac{1 + 4z}{3} \big[ (1 + 4 z) m_R + (1 + 2 z) m_p \big] (u - m_R^2)
\end{equation}
Non-pole contributions to the invariant amplitude $B$
\begin{equation} \label{eq:EQ046}
b_{u,np;2} = - \frac{1 + 4z}{3} 2 z (u - m_R^2)
\end{equation}

\subsubsection{\label{sec:RSu3}The contributions of the third term}

Pole contributions to the invariant amplitude $A$
\begin{equation} \label{eq:EQ047}
a_{u,p;3} = a_{s,p;3}
\end{equation}
Pole contributions to the invariant amplitude $B$
\begin{equation} \label{eq:EQ048}
b_{u,p;3} = - b_{s,p;3}
\end{equation}
Non-pole contributions to the invariant amplitude $A$
\begin{align} \label{eq:EQ049}
a_{u,np;3} = \frac{u - m_R^2}{3 m_R} \big[ & (1 + 8 z + 12 z^2) m_R^2 - (1 + 6 z + 6 z^2) m_p^2 + 2 z m_c^2 \nonumber \\
&+ (\frac{1}{2} + 4 z + 6 z^2) (u - m_R^2) \big]
\end{align}
Non-pole contributions to the invariant amplitude $B$
\begin{equation} \label{eq:EQ050}
b_{u,np;3} = \frac{u - m_R^2}{3 m_R} (1 + 6 z + 12 z^2) m_p
\end{equation}

\subsubsection{\label{sec:RSu4}The contributions of the fourth term}

Pole contributions to the invariant amplitude $A$
\begin{equation} \label{eq:EQ051}
a_{u,p;4} = a_{s,p;4}
\end{equation}
Pole contributions to the invariant amplitude $B$
\begin{equation} \label{eq:EQ052}
b_{u,p;4} = - b_{s,p;4}
\end{equation}
Non-pole contributions to the invariant amplitude $A$
\begin{align} \label{eq:EQ053}
a_{u,np;4} = - \frac{2 (u - m_R^2)}{3 m_R^2} \Big\{ & \frac{m_R + m_p}{2} (m_R^2 - m_p^2 + m_c^2) + z m_R (2 m_R^2 - 2 m_p^2 + m_c^2) \nonumber \\
&+ z^2 (m_R - m_p) (2 m_R^2 - m_p^2) \nonumber \\
&+ \big[ (\frac{1}{4} + z + z^2) m_R + (\frac{1}{4} - z^2) m_p \big] (u - m_R^2) \Big\}
\end{align}
Non-pole contributions to the invariant amplitude $B$
\begin{align} \label{eq:EQ054}
b_{u,np;4} = \frac{2 (u - m_R^2)}{3 m_R^2} \big[ & \frac{m_R^2 - m_p^2 + m_c^2}{2} + z (2 m_R^2 - m_R m_p - m_p^2 + m_c^2) \nonumber \\
&+ z^2 (2 m_R^2 - 2 m_R m_p + m_p^2) \nonumber \\
&+ (\frac{1}{4} + z + z^2) (u - m_R^2) \big]
\end{align}

\subsubsection{\label{sec:RSuTotal}Sums of the $u$-channel contributions}

The $u$-channel contributions to the invariant amplitudes $A$ and $B$ may be obtained, using the results detailed in Subsections \ref{sec:RSu1}-\ref{sec:RSu4}, via the expressions.
\begin{equation} \label{eq:EQ055}
A_{u,p} = \frac{1}{u - m_R^2} \left( \frac{g_{\pi N R}}{2 m_p} \right) ^2 \sum_{n=1}^4 a_{u,p;n}
\end{equation}
\begin{equation} \label{eq:EQ056}
B_{u,p} = \frac{1}{u - m_R^2} \left( \frac{g_{\pi N R}}{2 m_p} \right) ^2 \sum_{n=1}^4 b_{u,p;n}
\end{equation}
\begin{equation} \label{eq:EQ057}
A_{u,np} = \frac{1}{u - m_R^2} \left( \frac{g_{\pi N R}}{2 m_p} \right) ^2 \sum_{n=1}^4 a_{u,np;n}
\end{equation}
\begin{equation} \label{eq:EQ058}
B_{u,np} = \frac{1}{u - m_R^2} \left( \frac{g_{\pi N R}}{2 m_p} \right) ^2 \sum_{n=1}^4 b_{u,np;n}
\end{equation}

As $a_{u,p;n}=a_{s,p;n}$ in all cases,
\begin{equation} \label{eq:EQ059}
A_{u,p}^\pm = \binom{\alpha}{-\beta} \frac{g_{\pi N R}^2}{12 m_p^2} \frac{\alpha_1 + \alpha_2 t}{m_R^2 - u} \, \, \, ,
\end{equation}
where $\alpha_1$ and $\alpha_2$ are defined in Eqs.~(\ref{eq:EQ031}) and the isospin structure has now been included. In particular, for the $u$-channel graph with a $\Delta(1232)$ as intermediate state, one obtains:
\begin{equation} \label{eq:EQ060}
A_{u,p}^\pm = \binom{2}{-1} \frac{g_{\pi N \Delta}^2}{36 m_p^2} \frac{\alpha_1 + \alpha_2 t}{m_\Delta^2 - u} \, \, \, .
\end{equation}

As $b_{u,p;i}=-b_{s,p;i}$ in all cases,
\begin{equation} \label{eq:EQ061}
B_{u,p}^\pm = - \binom{\alpha}{-\beta} \frac{g_{\pi N R}^2}{12 m_p^2} \frac{\beta_1 + \beta_2 t}{m_R^2 - u} \, \, \, ,
\end{equation}
where $\beta_1$ and $\beta_2$ are defined in Eqs.~(\ref{eq:EQ035}). In particular, for the $u$-channel graph with a $\Delta(1232)$ as intermediate state, one obtains:
\begin{equation} \label{eq:EQ062}
B_{u,p}^\pm = - \binom{2}{-1} \frac{g_{\pi N \Delta}^2}{36 m_p^2} \frac{\beta_1 + \beta_2 t}{m_\Delta^2 - u} \, \, \, .
\end{equation}

The final expression for $A_{u,np}$ reads as:
\begin{align} \label{eq:EQ063}
A_{u,np}^\pm = \binom{\alpha}{-\beta} \frac{g_{\pi N R}^2}{12 m_p^2} \Big\{ &2 (-1 + z + 2 z^2) m_R + (-3 + 2 z^2) m_p - (1 + 2 z + 4 z^2) \frac{m_p^2}{m_R} \nonumber \\
&+ 2 p_{0R} + 2 \frac{m_p}{m_R} p_{0R} - 2 z^2 \frac{m_p^3}{m_R^2} \nonumber \\
&+ \frac{u - m_R^2}{m_R} \big[ 2 z + 4 z^2 - (\frac{1}{2} - 2 z^2) \frac{m_p}{m_R} \big] \Big\} \, \, \, .
\end{align}

The final expression for $B_{u,np}$ reads as:
\begin{align} \label{eq:EQ064}
B_{u,np}^\pm = \binom{\alpha}{-\beta} \frac{g_{\pi N R}^2}{12 m_p^2} \big[ &1 + 2z + 2z^2 + (1 + 4z + 8z^2) \frac{m_p}{m_R} \nonumber \\
&- (1 + 2z - 2z^2) \frac{m_p^2}{m_R^2} - 2 \frac{m_c^2}{m_R^2} Z + 2 \frac{u - m_R^2}{m_R^2} Z^2 \big] \, \, \, .
\end{align}

\subsection{\label{sec:RSTotal}Final expressions for the Rarita-Schwinger propagator}

The expressions for the pole contributions to the invariant amplitudes $A$ and $B$ are given by Eqs.~(\ref{eq:EQ030}) and (\ref{eq:EQ034}) for the $s$-channel graph, and by Eqs.~(\ref{eq:EQ059}) and (\ref{eq:EQ061}) for the 
$u$-channel graph. As earlier mentioned, inasmuch as they do not depend on the parameter $Z$, these contributions are unique. The sums of the $s$- and $u$-channel contributions are put in the concise forms:
\begin{equation} \label{eq:EQ064b}
A_{p}^\pm = \binom{\alpha}{\beta} \frac{g_{\pi N R}^2}{12 m_p^2} \left( \frac{\alpha_1 + \alpha_2 t}{m_R^2 - s} \pm \frac{\alpha_1 + \alpha_2 t}{m_R^2 - u} \right)
\end{equation}
and
\begin{equation} \label{eq:EQ064c}
B_{p}^\pm = \binom{\alpha}{\beta} \frac{g_{\pi N R}^2}{12 m_p^2} \left( \frac{\beta_1 + \beta_2 t}{m_R^2 - s} \mp \frac{\beta_1 + \beta_2 t}{m_R^2 - u} \right) \, \, \, .
\end{equation}

We now come to the expressions for the non-pole contributions to the invariant amplitudes $A$ and $B$.

Concerning the invariant amplitudes $A$, the sums of the $s$-channel of Eq.~(\ref{eq:EQ037}) and $u$-channel of Eq.~(\ref{eq:EQ063}) contributions read as:
\begin{align} \label{eq:EQ065}
A_{np}^+ = \alpha \frac{g_{\pi N R}^2}{12 m_p^2} \Big\{ &4 (-1 + z + 2 z^2) m_R + 2 (-3 + 2 z^2) m_p - 2 (1 + 2 z + 4 z^2) \frac{m_p^2}{m_R} \nonumber \\
&+ 4 p_{0R} + 4 \frac{m_p}{m_R} p_{0R} - 4 z^2 \frac{m_p^3}{m_R^2} \nonumber \\
&+ \frac{s + u - 2 m_R^2}{m_R} \big[ 2 z + 4 z^2 - (\frac{1}{2} - 2 z^2) \frac{m_p}{m_R} \big] \Big\}
\end{align}
and
\begin{equation} \label{eq:EQ066}
A_{np}^- = \beta \frac{g_{\pi N R}^2}{12 m_p^2 m_R} (s-u) \big[ 2 z + 4 z^2 - (\frac{1}{2} - 2 z^2) \frac{m_p}{m_R} \big] \, \, \, .
\end{equation}
To somewhat compactify the expressions, the variable $Y$ was introduced \cite{h} according to the equation:
\begin{equation} \label{eq:EQ067}
Y = (2 + \frac{m_p}{m_R}) Z^2 + (1 + \frac{m_p}{m_R}) Z = z + 2 z^2 - (\frac{1}{4} - z^2) \frac{m_p}{m_R} \, \, \, .
\end{equation}
Using Eq.~(\ref{eq:EQ067}), one may put Eqs.~(\ref{eq:EQ065}) and (\ref{eq:EQ066}) in the forms:
\begin{equation} \label{eq:EQ068}
A_{np}^+ = - \alpha \frac{g_{\pi N R}^2}{6 m_p^2 m_R} \big[ (p_{0R}+m_p) (2 m_R-m_p) + (2 + \frac{m_p}{2 m_R}) m_c^2 + (t - 2 m_c^2) Y \big]
\end{equation}
and
\begin{equation} \label{eq:EQ069}
A_{np}^- = \beta \frac{2 g_{\pi N R}^2}{3 m_p m_R} Y \nu \, \, \, .
\end{equation}
The Mandelstam variable $\nu$ has been defined in Eq.~(\ref{eq:EQ000}).

Regarding the invariant amplitudes $B$, the sums of the $s$-channel of Eq.~(\ref{eq:EQ038}) and $u$-channel of Eq.~(\ref{eq:EQ064}) contributions lead to:
\begin{equation} \label{eq:EQ070}
B_{np}^+ = - \alpha \frac{2 g_{\pi N R}^2}{3 m_p m_R^2} Z^2 \nu
\end{equation}
and
\begin{equation} \label{eq:EQ072}
B_{np}^- = - \beta \frac{g_{\pi N R}^2}{12 m_p^2} \Big\{ (1+\frac{m_p}{m_R})^2 +\frac{8m_pY}{m_R} +\frac{4}{m_R^2} \big[ (m_c^2 -\frac{t}{2}) Z^2 - m_c^2 Z \big] \Big\} \, \, \, .
\end{equation}

The final expressions of this subsection agree with the formulae of Refs.~\cite{h,nek}, which (up to the present time) had been used in the ETH model of the $\pi N$ interaction \cite{glm,mr,glmbg} without verification. In this 
paper, we gave all the contributions in a form which facilitates their use in the general case of isospin decomposition of the scattering amplitude. It is now easy to understand the somewhat peculiar factors which had been 
applied to the pole and non-pole contributions of the $\Delta$ amplitudes in Ref.~\cite{mr} (see Subsection 3.5.3 therein), in order to determine the contributions of the graphs with an $N(1720)$ intermediate state. As an example, 
in order to obtain the conversion factor for the non-pole isovector part of the invariant amplitudes $A$ and $B$, one must first `undo' the isospin decomposition of the $\Delta(1232)$ in that part; this implies a multiplication by 
$3$ (as the factor originally applied had been $\beta=\frac{1}{3}$. Subsequently, one must multiply by $-1$, i.e., the appropriate $\beta$ value for an $N$-type isospin decomposition. Therefore, the overall factor is equal to 
$3 \cdot (-1)=-3$.

To summarise, the pole contributions to the invariant amplitudes $A$ are given in Eq.~(\ref{eq:EQ064b}); the pole contributions to the invariant amplitudes $B$ are given in Eq.~(\ref{eq:EQ064c}). The non-pole contributions to the 
invariant amplitudes $A$ are given by Eqs.~(\ref{eq:EQ068}) and (\ref{eq:EQ069}). Finally, the non-pole contributions to the invariant amplitudes $B$ are given by Eqs.~(\ref{eq:EQ070}) and (\ref{eq:EQ072}). The pole contributions 
are unique, in that they do not depend on the parameter $Z$; the non-pole contributions are $Z$-dependent. Shown in Figs.~\ref{fig:RS}, as functions of the Mandelstam variables $s$ and $t$, are the contributions to the invariant 
amplitudes $A$ and $B$ of the $\Delta(1232)$ graphs (Figs.~\ref{fig:PiNDelta}), obtained with the Rarita-Schwinger propagator. The partial-wave decomposition of the scattering amplitude, obtained with the Rarita-Schwinger 
propagator for the $\Delta(1232)$ graphs, has been given in Subsection 3.4 of Ref.~\cite{mr}. The appropriate isospin decomposition of the scattering amplitude is obtained by suitably choosing the values of the quantities $\alpha$ 
and $\beta$, fixing the isoscalar and isovector contributions, respectively.

The expressions of the present section are drastically simplified when $z=0$ or, equivalently, $Z=-\frac{1}{2}$. In this context, it is interesting to mention that, for a long time, the results of the fits to meson-factory 
low-energy $\pi^\pm p$ elastic-scattering data (see Ref.~\cite{mr} and the relevant references therein) have been compatible~\footnote{The relative weakness of the contributions to the invariant amplitudes of the terms which are 
linear in $Z$, in conjunction with the significant correlations which are present among the model parameters during the optimisation (especially when the floating of the experimental data sets is allowed), leads to a weak dependence 
of the $\chi^2$ value on the sign of the parameter $Z$; the analysis of the results of the fits at fixed $Z$ values demonstrated that the $\chi^2$ function is nearly symmetric around $Z=0$, creating two deep local minima, one around 
$Z=\frac{1}{2}$, the other around $Z=-\frac{1}{2}$. Fortunately enough, up to now, the minimum around $Z=-\frac{1}{2}$ has always been deeper and, more importantly, the one at positive $Z$ yields an unphysical result for the model 
parameter $\kappa_\rho$; $\kappa_\rho$ is negative for $Z \gtrsim 0.15$, reaching the value of about $-4.5$ at $Z=\frac{1}{2}$.} with the solution $Z=-\frac{1}{2}$. This remark may be useful in the calculation of the contributions 
to the scattering amplitude of graphs with massive intermediate states of higher spin ($J > \frac{3}{2}$).

\section{\label{sec:WL}The Williams propagator}

The operator $\mathscr{P}_{\mu \nu} (P)$, corresponding to the Williams propagator, is:
\begin{equation} \label{eq:EQ073}
\mathscr{P}_{\mu \nu}^{WL} (P) = \mathscr{P}_{\mu \nu}^{3/2} (P) = g_{\mu \nu} - \frac{1}{3} \gamma_\mu \gamma_\nu - \frac{\fsl{P} \gamma_\mu P_\nu + \gamma_\nu P_\mu \fsl{P}}{3 P^2} \, \, \, ,
\end{equation}
where $\mathscr{P}_{\mu \nu}^{3/2} (P)$ has been introduced in Eq.~(\ref{eq:EQ008_1}). It satisfies
\begin{equation} \label{eq:EQ074}
\gamma^\mu \mathscr{P}_{\mu \nu}^{WL} (P) = \mathscr{P}_{\mu \nu}^{WL} (P) \gamma^\nu = 0
\end{equation}
and
\begin{align} \label{eq:EQ075}
P^\mu \mathscr{P}_{\mu \nu}^{WL} (P) = P^\nu \mathscr{P}_{\mu \nu}^{WL} (P) = 0 \, \, \, .
\end{align}

On the whole, the calculations of the contributions to the scattering amplitude with the Williams propagator are significantly shorter (and less tedious) than in the Rarita-Schwinger case.

\subsection{\label{sec:WLs}The $s$-channel contributions}

Once again, explicit results will be given for each of the four terms comprising the operator of Eq.~(\ref{eq:EQ073}). For brevity, these contributions will be split into pole and non-pole parts at a later stage, after listing the 
$u$-channel results. In the contributions of Subsections \ref{sec:WLs1}-\ref{sec:WLs4}, the factor $\frac{1}{s - m_R^2} \left( \frac{g_{\pi N R}}{2 m_p} \right) ^2$, appearing on the rhs of Eq.~(\ref{eq:EQ006}), is suppressed.

\subsubsection{\label{sec:WLs1}The contributions of the first term}

Contributions to the invariant amplitude $A$
\begin{equation} \label{eq:EQ076}
a_{s;1} = (m_R + m_p) (m_c^2 - \frac{t}{2})
\end{equation}
Contributions to the invariant amplitude $B$
\begin{equation} \label{eq:EQ077}
b_{s;1} = m_c^2 - \frac{t}{2}
\end{equation}

\subsubsection{\label{sec:WLs2}The contributions of the second term}

Contributions to the invariant amplitude $A$
\begin{equation} \label{eq:EQ078}
a_{s;2} = - \frac{1}{3} (m_R + m_p) (s - m_p^2)
\end{equation}
Contributions to the invariant amplitude $B$
\begin{equation} \label{eq:EQ079}
b_{s;2} = \frac{1}{3} (2 m_R m_p + 2 m_p^2 - m_c^2)
\end{equation}

\subsubsection{\label{sec:WLs3}The contributions of the third term}

Contributions to the invariant amplitude $A$
\begin{equation} \label{eq:EQ080}
a_{s;3} = - \frac{s - m_p^2 + m_c^2}{6s} m_R m_c^2
\end{equation}
Contributions to the invariant amplitude $B$
\begin{equation} \label{eq:EQ081}
b_{s;3} = - \frac{s - m_p^2 + m_c^2}{6s} (s + m_R m_p)
\end{equation}

\subsubsection{\label{sec:WLs4}The contributions of the fourth term}

Contributions to the invariant amplitude $A$
\begin{equation} \label{eq:EQ082}
a_{s;4} = - \frac{s - m_p^2 + m_c^2}{6s} \big[ (m_R + m_p) m_c^2 + m_p (s - m_p^2) \big]
\end{equation}
Contributions to the invariant amplitude $B$
\begin{equation} \label{eq:EQ083}
b_{s;4} = - \frac{s - m_p^2 + m_c^2}{6s} (m_R m_p - m_p^2 + m_c^2)
\end{equation}

\subsection{\label{sec:WLu}The $u$-channel contributions}

In the contributions of Subsections \ref{sec:WLu1}-\ref{sec:WLu4}, the factor $\frac{1}{u - m_R^2} \left( \frac{g_{\pi N R}}{2 m_p} \right) ^2$, appearing on the rhs of Eq.~(\ref{eq:EQ007}), is suppressed.

\subsubsection{\label{sec:WLu1}The contributions of the first term}

Contributions to the invariant amplitude $A$
\begin{equation} \label{eq:EQ084}
a_{u;1} = a_{s;1}
\end{equation}
Contributions to the invariant amplitude $B$
\begin{equation} \label{eq:EQ085}
b_{u;1} = - b_{s;1}
\end{equation}

\subsubsection{\label{sec:WLu2}The contributions of the second term}

Contributions to the invariant amplitude $A$
\begin{equation} \label{eq:EQ086}
a_{u;2} = - \frac{1}{3} (m_R + m_p) (u - m_p^2)
\end{equation}
Contributions to the invariant amplitude $B$
\begin{equation} \label{eq:EQ087}
b_{u;2} = - b_{s;2}
\end{equation}

\subsubsection{\label{sec:WLu3}The contributions of the third term}

Contributions to the invariant amplitude $A$
\begin{equation} \label{eq:EQ088}
a_{u;3} = - \frac{u - m_p^2 + m_c^2}{6u} m_R m_c^2
\end{equation}
Contributions to the invariant amplitude $B$
\begin{equation} \label{eq:EQ089}
b_{u;3} = \frac{u - m_p^2 + m_c^2}{6u} (u + m_R m_p)
\end{equation}

\subsubsection{\label{sec:WLu4}The contributions of the fourth term}

Contributions to the invariant amplitude $A$
\begin{equation} \label{eq:EQ090}
a_{u;4} = - \frac{u - m_p^2 + m_c^2}{6u} \big[ (m_R + m_p) m_c^2 + m_p (u - m_p^2) \big]
\end{equation}
Contributions to the invariant amplitude $B$
\begin{equation} \label{eq:EQ091}
b_{u;4} = \frac{u - m_p^2 + m_c^2}{6u} (m_R m_p - m_p^2 + m_c^2)
\end{equation}

\subsection{\label{sec:WLTotal}Final expressions for the Williams propagator}

The pole contributions to the invariant amplitude $A$ may be put in the convenient form
\begin{equation} \label{eq:EQ092}
A_{p}^\pm = \binom{\alpha}{\beta} \frac{g_{\pi N R}^2}{12 m_p^2} \left( \frac{\alpha_1^\prime + \alpha_2 t}{m_R^2 - s} \pm \frac{\alpha_1^{\prime \prime} + \alpha_2 t}{m_R^2 - u} \right) \, \, \, ,
\end{equation}
where the coefficient $\alpha_2$ is defined in the second of Eqs.~(\ref{eq:EQ031}); $\alpha_1^\prime=m_R^2 \alpha_1 / s$ and $\alpha_1^{\prime \prime}=m_R^2 \alpha_1 / u$, where $\alpha_1$ is defined in the first of 
Eqs.~(\ref{eq:EQ031}).

Similarly,
\begin{equation} \label{eq:EQ093}
B_{p}^\pm = \binom{\alpha}{\beta} \frac{g_{\pi N R}^2}{12 m_p^2} \left( \frac{\beta_1^\prime + \beta_2 t}{m_R^2 - s} \mp \frac{\beta_1^{\prime \prime} + \beta_2 t}{m_R^2 - u} \right) \, \, \, ,
\end{equation}
where the coefficient $\beta_2$ is defined in the second of Eqs.~(\ref{eq:EQ035}); $\beta_1^\prime=m_R^2 \beta_1 / s$ and $\beta_1^{\prime \prime}=m_R^2 \beta_1 / u$, where $\beta_1$ is defined in the first of Eqs.~(\ref{eq:EQ035}).

The final expressions for the non-pole contributions to the invariant amplitudes $A$ read as:
\begin{equation} \label{eq:EQ094}
A_{np}^+ = - \alpha \frac{g_{\pi N R}^2}{24 m_p^2} \big[ 4 m_R + 6 m_p + \lambda_1 (\frac{1}{s} + \frac{1}{u}) \big]
\end{equation}
and
\begin{equation} \label{eq:EQ095}
A_{np}^- = \beta \lambda_1 \frac{g_{\pi N R}^2}{6 m_p s u} \nu \, \, \, ,
\end{equation}
where $\lambda_1=m_R^2 (2 m_R + m_p) + 2 (m_R + m_p) (m_R m_p - 2 m_p^2 - 2 m_c^2)$.

The final expressions for the non-pole contributions to the invariant amplitudes $B$ read as:
\begin{equation} \label{eq:EQ096}
B_{np}^+ = - \alpha \lambda_2 \frac{g_{\pi N R}^2}{6 m_p s u} \nu \, \, \, ,
\end{equation}
and
\begin{equation} \label{eq:EQ097}
B_{np}^- = \beta \frac{g_{\pi N R}^2}{24 m_p^2} \big[ -2 + \lambda_2 (\frac{1}{s} + \frac{1}{u}) \big] \, \, \, ,
\end{equation}
where $\lambda_2=- m_R^2 + 2 m_R m_p + 6 m_p^2 + 2 m_c^2$.

The partial-wave amplitudes obtained with the Williams propagator may be found in Appendix \ref{App:AppB}.

\section{\label{sec:PL}Pascalutsa's method}

In Refs.~\cite{pasca1,pasca2}, Pascalutsa argued that the vertex factor of Eq.~(\ref{eq:EQ002}) (in his papers, the `inconsistent coupling') must be corrected in order to obey the physical degrees of freedom counting, thus leading 
to what he considered to be the `consistent coupling'. In Ref.~\cite{pasca2}, he demonstrated that these two types are related via a redefinition of the massive spin-$\frac{3}{2}$ field. From the practical point of view, the use 
of the `consistent coupling' along with the operator $\mathscr{P}_{\mu \nu}^{3/2} (P)$ of Eq.~(\ref{eq:EQ073}) is mathematically equivalent to the use of the `inconsistent coupling' along with the multiplication of 
$\mathscr{P}_{\mu \nu}^{3/2} (P)$ by $\frac{P^2}{m_R^2}$. Owing to Eqs.~(\ref{eq:EQ074}) and (\ref{eq:EQ075}), the contributions to the scattering amplitude are not $Z$-dependent.

The extraction of the contributions to the invariant amplitudes in Pascalutsa's method may be easily obtained from the results of Subsections \ref{sec:WLs} and \ref{sec:WLu} for the Williams propagator, after the rhs of all the 
expressions of Subsection \ref{sec:WLs} is multiplied by $\frac{s}{m_R^2}$ and of all those of Subsection \ref{sec:WLu} by $\frac{u}{m_R^2}$. The derivation of the expressions is straightforward; we directly come to the results.

The pole part of the invariant amplitudes $A$ and $B$ is identical to the Rarita-Schwinger forms, given by Eqs.~(\ref{eq:EQ064b}) and (\ref{eq:EQ064c}), respectively. The partial-wave decomposition of the pole part of the 
scattering amplitude for the Rarita-Schwinger propagator may be found in Ref.~\cite{mr}.

The final expressions for the non-pole contributions to the invariant amplitudes $A$ read as:
\begin{equation} \label{eq:EQ098}
A_{np}^+ = - \alpha \frac{g_{\pi N R}^2}{24 m_p^2 m_R^2} \big[ (4 m_R + 3 m_p) t + 4 m_R (2 m_R + m_p) p_{0R} + 4 m_p (m_R^2 - m_R m_p - m_p^2) \big]
\end{equation}
and
\begin{equation} \label{eq:EQ099}
A_{np}^- = - \beta \frac{g_{\pi N R}^2}{6 m_p m_R^2} (2 m_R + 3 m_p) \nu \, \, \, .
\end{equation}

The final expressions for the non-pole contributions to the invariant amplitudes $B$ read as:
\begin{equation} \label{eq:EQ100}
B_{np}^+ = - \alpha \frac{g_{\pi N R}^2}{6 m_p m_R^2} \nu \, \, \, ,
\end{equation}
and
\begin{equation} \label{eq:EQ101}
B_{np}^- = \beta \frac{g_{\pi N R}^2}{24 m_p^2 m_R^2} ( -5 t + 4 m_R m_p - 4 m_R p_{0R} + 12 m_p^2) \, \, \, .
\end{equation}

One can show that the isoscalar part of the $\pi N$ scattering amplitude, obtained in Pascalutsa's method, is identical to the one obtained with the Rarita-Schwinger propagator and $Z=\frac{1}{2}$.

The partial-wave amplitudes (non-pole contributions), obtained in Pascalutsa's method, may be found in Appendix \ref{App:AppC}.

\section{\label{sec:Comparison}Comparison of results obtained in the three approaches}

In the present section, we will compare the invariant amplitudes $A$ and $B$ of the $\Delta(1232)$ graphs (Figs.~\ref{fig:PiNDelta}), obtained in Sections \ref{sec:RS}, \ref{sec:WL}, and \ref{sec:PL}. This comparison will involve 
the same fixed value of the coupling constant $g_{\pi N \Delta}$, taken from a fit of the ETH model to low-energy $\pi^\pm p$ elastic-scattering data \cite{mr}; the value of the parameter $Z$, entering the Rarita-Schwinger amplitudes, 
will be taken from the same fit. The fitted $g_{\pi N \Delta}$ and $Z$ results for $m_\sigma=475$ MeV (i.e., for the central value of the recommended range for the $\sigma$-meson mass \cite{pdg}, see Ref.~\cite{mr} for details) 
are $29.81$ and $-0.565$, respectively. All other physical constants needed in the comparison have been fixed from Ref.~\cite{pdg}. Of course, a more meaningful comparison should rather involve the fitted values of the coupling 
constant $g_{\pi N \Delta}$, obtained from separate fits to the low-energy data of Ref.~\cite{mr}, using the Williams propagator and Pascalutsa's method (instead of the Rarita-Schwinger propagator); however, this work is left for 
the future.

Using as reference the invariant amplitudes $A$ and $B$ of the $\Delta(1232)$ graphs obtained with the Rarita-Schwinger propagator (shown in Figs.~\ref{fig:RS}), we estimated the limits of the relative differences in the 
low-energy region ($T \leq 100$ MeV) for the other two treatments investigated in the present paper; these limits are shown in Table \ref{tab:DiffToRS}. Each difference is defined as $f(s,t)/f_{RS}(s,t)-1$, where $f(s,t)$ stands 
for the value of one amplitude (i.e., of $A^+$, $A^-$, $B^+$, or $B^-$), obtained with the Williams propagator or in Pascalutsa's approach, at the specific ($s$,$t$) point, and $f_{RS}(s,t)$ is the corresponding result obtained 
with the Rarita-Schwinger propagator. The largest relative differences in Table \ref{tab:DiffToRS} correspond to the isovector part of the $\pi N$ scattering amplitude obtained in Pascalutsa's method. The absolute differences 
$\delta f(s,t)=f(s,t)-f_{RS}(s,t)$ are shown in Figs.~\ref{fig:DiffToRS}.

The contributions to the $s$-wave scattering lengths, obtained in the Williams and Pascalutsa approaches, vanish. Within the Rarita-Schwinger formalism, null contributions to both scattering lengths can be obtained only when 
$Z=\frac{1}{2}$; the corresponding expressions for the contributions to the isoscalar ($b_0$) and isovector ($b_1$) scattering lengths are:
\begin{equation} \label{eq:EQ101b}
b_0 = - \alpha \frac{g_{\pi N R}^2 \, m_c^2}{12 \pi m_R (m_p + m_c)} \left( Z - \frac{1}{2} \right) \left( \frac{Z-\frac{1}{2}}{m_R}-\frac{2(Z+1)}{m_p} \right)
\end{equation}
and
\begin{equation} \label{eq:EQ101c}
b_1 = - \beta \frac{g_{\pi N R}^2 \, m_c^3}{12 \pi m_p m_R^2 (m_p + m_c)} \left( Z - \frac{1}{2} \right) ^2 \, \, \, .
\end{equation}
In view of the fact that, in our PWAs of the low-energy $\pi^\pm p$ elastic-scattering data, the fitted values of the parameter $Z$ come out close to $Z=-\frac{1}{2}$ (i.e., signifying sizeable contributions to the $s$ waves), the 
reproduction of the experimental data on the basis of the Williams and Pascalutsa approaches merits an investigation; however, such an analysis goes beyond the narrow score of the present paper.

It is worth mentioning that the contributions of the graphs, involving a massive spin-$\frac{3}{2}$ intermediate state, to the $\pi N$ $\Sigma$ term are non-zero in the three approaches investigated herein. Using Eq.~(\ref{eq:EQ064b}) 
to obtain $A_{p}^+$ and Eq.~(\ref{eq:EQ068}) to obtain $A_{np}^+$, summing up the pole and non-pole contributions at the Cheng-Dashen point ($s$,$t$)=($m_p^2$,$2 m_c^2$), and multiplying the result by the pion-decay constant $F_\pi$, 
one obtains for the Rarita-Schwinger propagator:
\begin{equation} \label{eq:EQ102}
\Sigma^{RS} = \alpha F_\pi^2 \frac{g_{\pi N R}^2 \, (2 m_R + m_p) \, m_c^4}{12 m_p^2 m_R^2 (m_R^2 - m_p^2)} \, \, \, ;
\end{equation}
$\Sigma^{RS}$ turns out to be independent of $Z$. The same result is obtained in Pascalutsa's method. The result obtained with the Williams propagator is:
\begin{equation} \label{eq:EQ103}
\Sigma^{WL} = \alpha F_\pi^2 \frac{g_{\pi N R}^2 \, (2 m_R + m_p) \, m_c^4}{12 m_p^4 (m_R^2 - m_p^2)} \, \, \, .
\end{equation}

\section{\label{sec:Conclusions}Summary}

In the present paper, we set on investigating the propagation of a massive spin-$\frac{3}{2}$ intermediate state, aiming at a direct application in hadronic models of the pion-nucleon ($\pi N$) interaction. To facilitate the use 
of our results, suitable expressions have been given, applicable in the general case of isospin decomposition of the scattering amplitude; this is enabled via the use of two quantities in the expressions, $\alpha$ (pertaining to 
the isoscalar contributions) and $\beta$ (pertaining to the isovector contributions).

In Section \ref{sec:RS}, we dealt with the details of the lengthy derivation of the contributions to the standard invariant amplitudes $A$ and $B$ using the Rarita-Schwinger propagator; the final results for this propagator have 
been known since a long time, yet a) the details of the calculation had not appeared in the original papers and b) we are not aware of any attempts to verify the validity of the expressions found in the literature. After repeating 
the calculation, we confirmed the expressions for the amplitudes given in Nath \etal \cite{nek} and H\"ohler \cite{h}.

In Sections \ref{sec:WL} and \ref{sec:PL}, we derived the contributions to the invariant amplitudes $A$ and $B$ following two other approaches, namely using the Williams propagator \cite{wl}, which was introduced in the mid 1980s, 
and the method which Pascalutsa \cite{pasca1,pasca2} proposed more recently. Detailed expressions for these contributions are given in a form which can easily be used in other works.

\begin{figure}
\begin{center}
\includegraphics [width=15.5cm] {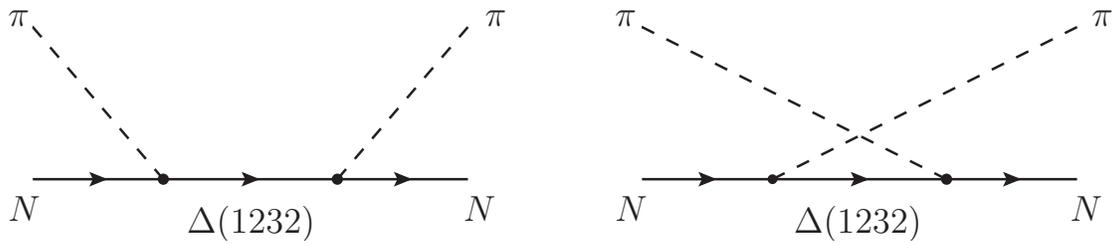}
\caption{\label{fig:PiNDelta}The $\Delta(1232)$ $s$- and $u$-channel graphs, used in hadronic models of the $\pi N$ scattering. Implemented in the ETH model \cite{glm,mr,glmbg} (but not shown here) are also the graphs corresponding 
to another massive spin-$\frac{3}{2}$ state, namely to the $N(1720)$.}
\end{center}
\end{figure}

\begin{figure}
\begin{center}
\includegraphics [width=15.5cm] {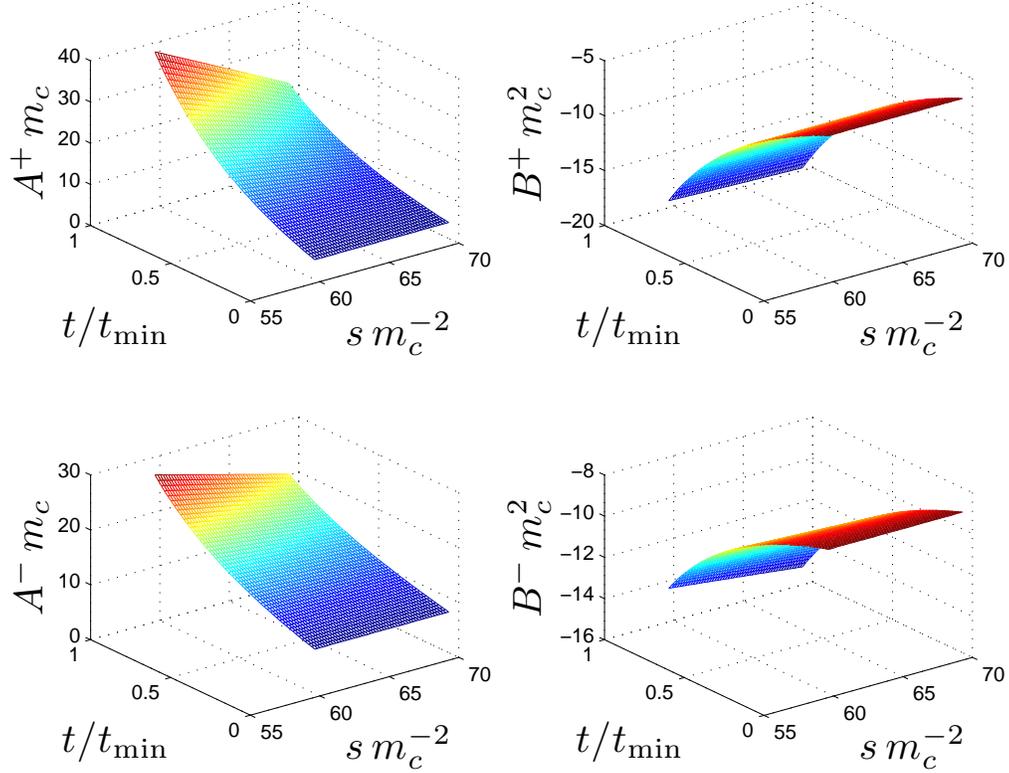}
\caption{\label{fig:RS}The sum of the pole and non-pole contributions to the invariant amplitudes $A$ and $B$ of the $\Delta(1232)$ graphs (Figs.~\ref{fig:PiNDelta}) using the Rarita-Schwinger propagator (Section \ref{sec:RS}). 
The values of $g_{\pi N \Delta}$ and $Z$ have been obtained from a recent fit of the ETH model to low-energy $\pi^\pm p$ elastic-scattering data \cite{mr} (see Section \ref{sec:Conclusions}). The invariant amplitudes are shown as 
functions of the two independent Mandelstam variables $s$ and $t$. The ($s$-dependent) value of $t_{\rm min}$ is equal to $-4 \vec{q} \, ^2$. The smallest $s$ value represents the $\pi N$ threshold, i.e., pion laboratory kinetic 
energy $T=0$ MeV; the largest $s$ value corresponds to $T=100$ MeV.}
\end{center}
\end{figure}

\begin{figure}
\begin{center}
\includegraphics [width=15.5cm] {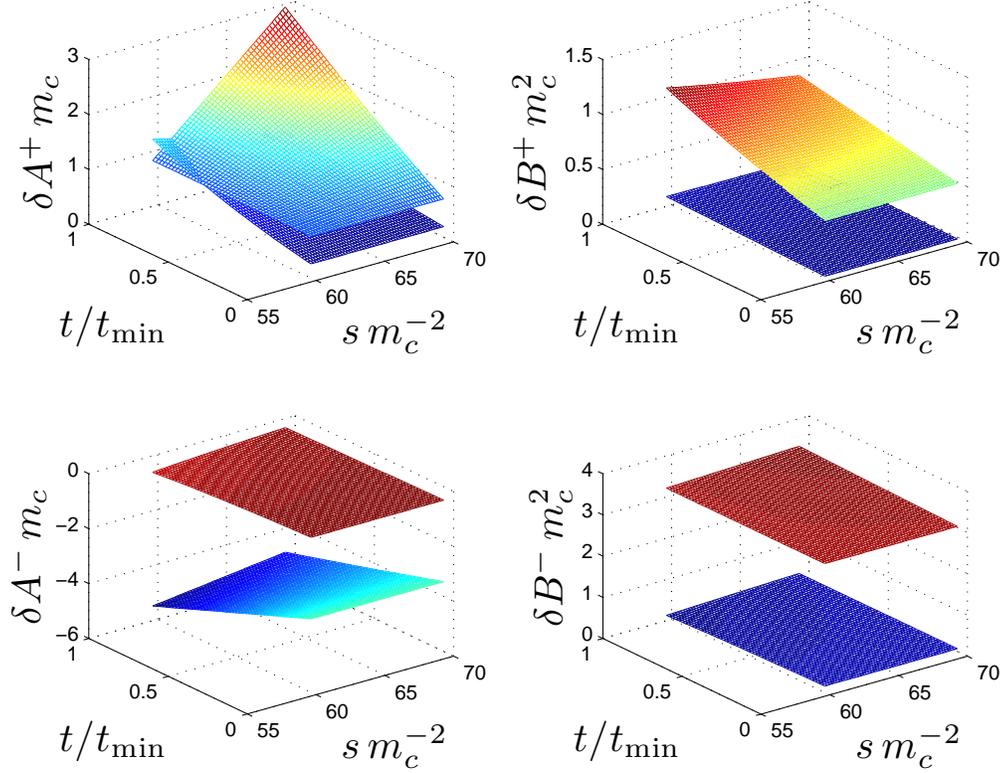}
\caption{\label{fig:DiffToRS}The absolute differences of the values of the invariant amplitudes $A$ and $B$ of the $\Delta(1232)$ graphs (Figs.~\ref{fig:PiNDelta}), obtained with the Williams propagator and in Pascalutsa's method, 
to the amplitudes obtained with the Rarita-Schwinger propagator. The values of $g_{\pi N \Delta}$ and $Z$ have been obtained from a recent fit of the ETH model to low-energy $\pi^\pm p$ elastic-scattering data \cite{mr} (see 
Section \ref{sec:Conclusions}). The differences are shown as functions of the two independent Mandelstam variables $s$ and $t$. The ($s$-dependent) value of $t_{\rm min}$ is equal to $-4 \vec{q} \, ^2$. The smallest $s$ value 
represents the $\pi N$ threshold, i.e., pion laboratory kinetic energy $T=0$ MeV; the largest $s$ value corresponds to $T=100$ MeV. The surface corresponding to Pascalutsa's method is above the one obtained with the Williams 
propagator in $\delta A^+$ and $\delta B^-$, below in the two other cases.}
\end{center}
\end{figure}

In Section \ref{sec:Comparison}, the results obtained in the low-energy region for the invariant amplitudes $A$ and $B$ of the $\Delta(1232)$ graphs (Figs.~\ref{fig:PiNDelta}), in the three approaches investigated herein, are 
compared at fixed $g_{\pi N \Delta}$ (Table \ref{tab:DiffToRS} and Figs.~\ref{fig:DiffToRS}). Relative to the Rarita-Schwinger results, the largest differences observed pertain to the isovector part of the $\pi N$ scattering 
amplitude in Pascalutsa's method. Finally, we give analytical expressions for the $s$-wave scattering lengths, as well as for the $\pi N$ $\Sigma$ term in the three treatments.

\begin{ack}
One of us (E.M.) acknowledges a useful exchange of electronic mail with V. Pascalutsa.

The Feynman graphs of the present document have been drawn with the software package JaxoDraw \cite{JaxoDraw}, available from the link: http://jaxodraw.sourceforge.net/.
\end{ack}

\newpage
\begin{table}[h!]
{\bf \caption{\label{tab:DiffToRS}}}The limits in the differences of the values of the invariant amplitudes $A$ and $B$ of the $\Delta(1232)$ graphs (Figs.~\ref{fig:PiNDelta}) in the low-energy region (pion laboratory kinetic 
energy $T \leq 100$ MeV), obtained with the Williams propagator (WL) and in Pascalutsa's method (PL), relative (and normalised) to the results obtained with the Rarita-Schwinger propagator. The values of $g_{\pi N \Delta}$ and 
$Z$ have been obtained from a recent fit of the ETH model to low-energy $\pi^\pm p$ elastic-scattering data \cite{mr} (see Section \ref{sec:Conclusions}). The values have been truncated to one decimal digit.
\vspace{0.2cm}
\begin{center}
\begin{tabular}{|c|c|c|}
\hline
Invariant amplitude & WL & PL \\
\hline
$A^+$ & $ 3.1$ to $ 5.8 \%$ & $  2.2$ to $ 17.9 \%$ \\
$B^+$ & $-8.5$ to $-4.3 \%$ & $ -0.8$ to $ -0.4 \%$ \\
$A^-$ & $-2.5$ to $-1.6 \%$ & $-36.3$ to $-20.4 \%$ \\
$B^-$ & $-2.8$ to $-0.9 \%$ & $-36.5$ to $-22.0 \%$ \\
\hline
\end{tabular}
\end{center}
\end{table}

\clearpage

\clearpage

\clearpage

\clearpage
\newpage
\appendix
\section{\label{App:AppA}Useful relations}

Given below are a few important relations for the evaluation of the various contributions to the $T$-matrix elements.

\begin{equation*}
s + u + t = 2 m_p^2 + 2 m_c^2
\end{equation*}
\begin{equation*}
q^\prime \cdot q = q \cdot q^\prime = \frac{2 m_c^2 - t}{2}
\end{equation*}
\begin{equation*}
q^\prime \cdot p = p \cdot q^\prime = \frac{m_p^2 + m_c^2 - u}{2}
\end{equation*}
\begin{equation*}
p \cdot q = q \cdot p = \frac{s - m_p^2 - m_c^2}{2}
\end{equation*}
\begin{equation*}
\fsl{q} \fsl{q} = \fsl{q}^{\, \prime} \fsl{q}^{\, \prime} = m_c^2
\end{equation*}

In the correspondences below, the operators are assumed sandwiched between $\bar{u}_f (p^\prime)$ and $u_i (p)$. These expressions are useful in the evaluation of the contributions given in Sections \ref{sec:RS}, \ref{sec:WL}, 
and \ref{sec:PL}.

\begin{equation*}
\fsl{q} \rightarrow \gamma^0 W - m_p
\end{equation*}
\begin{equation*}
\fsl{q}^{\, \prime} \rightarrow \gamma^0 W - m_p
\end{equation*}
\begin{equation*}
\fsl{q}^{\, \prime} \fsl{q} \rightarrow s - m_p^2 - 2 m_p (\gamma^0 W - m_p)
\end{equation*}
\begin{equation*}
\fsl{p} \fsl{q}^{\, \prime} \rightarrow m_p^2 + m_c^2 - u - m_p (\gamma^0 W - m_p)
\end{equation*}
\begin{equation*}
\fsl{q} \fsl{q}^{\, \prime} \rightarrow u - m_p^2 + 2 m_p (\gamma^0 W - m_p)
\end{equation*}
\begin{equation*}
\fsl{p} \fsl{q} \rightarrow s - m_p^2 - m_c^2 - m_p (\gamma^0 W - m_p)
\end{equation*}
\begin{equation*}
\fsl{q} \fsl{q}^{\, \prime} \fsl{q} \rightarrow (m_c^2 - t) (\gamma^0 W - m_p)
\end{equation*}
\begin{equation*}
\fsl{q}^{\, \prime} \fsl{q} \fsl{q}^{\, \prime} \rightarrow (m_c^2 - t) (\gamma^0 W - m_p)
\end{equation*}
\begin{equation*}
\fsl{p} \fsl{q}^{\, \prime} \fsl{q} \rightarrow m_p (s - m_p^2) + (t - 2 m_p^2) (\gamma^0 W - m_p)
\end{equation*}
\begin{equation*}
\fsl{p} \fsl{q} \fsl{q}^{\, \prime} \rightarrow m_p (u - m_p^2) - (t - 2 m_p^2) (\gamma^0 W - m_p)
\end{equation*}
\begin{equation*}
\fsl{q}^{\, \prime} \fsl{p} \fsl{q} \rightarrow - m_p (s - m_p^2) + (s + m_p^2 - m_c^2) (\gamma^0 W - m_p)
\end{equation*}
\begin{equation*}
\fsl{q} \fsl{p} \fsl{q}^{\, \prime} \rightarrow - m_p (u - m_p^2) - (u + m_p^2 - m_c^2) (\gamma^0 W - m_p)
\end{equation*}
\begin{equation*}
\fsl{p} \fsl{q} \fsl{p} \fsl{q}^{\, \prime} \rightarrow (s - m_c^2) (m_p^2 + m_c^2 - u) - m_p^2 m_c^2 - (s + m_p^2 - m_c^2) m_p (\gamma^0 W - m_p)
\end{equation*}
\begin{equation*}
\fsl{q}^{\, \prime} \fsl{q} \fsl{p} \fsl{q}^{\, \prime} \rightarrow (m_p^2 + m_c^2 - u) (s - m_p^2) + (u - s - m_c^2) m_p (\gamma^0 W - m_p)
\end{equation*}

\clearpage
\newpage
\section{\label{App:AppB}Partial-wave decomposition of the scattering amplitude obtained with the Williams propagator}

Regarding the partial-wave decomposition of the scattering amplitude, the details are given in Ref.~\cite{mr}. Integrals of the form
\begin{equation*}
\phi_n (a,b) = \int_{-1}^{1} \frac{\xi^n \, d\xi}{a+b\xi} 
\end{equation*}
(where $\lvert a \rvert > \lvert b \rvert$) appear repeatedly in the partial-wave decomposition of the hadronic part of the scattering amplitude. For $b \neq 0$,
\begin{align*}
\phi_0 (a,b) &= \frac{1}{b} \ln \frac{a+b}{a-b} \, \, \, , \qquad \phi_1 (a,b) = \frac{2 - a \, \phi_0 (a,b)}{b} \, \, \, , \nonumber \\
\phi_2 (a,b) &= - \frac{a}{b} \phi_1 (a,b) \, \, \, , \qquad \phi_3 (a,b) = \frac{2/3 - a \, \phi_2 (a,b)}{b} \, \, \, , \nonumber \\
\phi_4 (a,b) &= - \frac{a}{b} \phi_3 (a,b) \, \, \, , \qquad \phi_5 (a,b) = \frac{2/5 - a \, \phi_4 (a,b)}{b} \, \, \, \dots
\end{align*}
Herein, we need the $\phi_n (a,b)$ for $n \leq 4$.

\subsection{\label{sec:Pole}Pole contributions}

The $s$-channel graph leads to the expressions.
\begin{align*}
K_{0+} &= \lambda \big( (\alpha_0 + \beta_-) (p_0 + m_p) + \frac{2 \vec{q} \, ^2}{3} (-\alpha_2 + \beta_+^\prime) (p_0 - m_p) \big) \nonumber \\
K_{1+} &= \frac{2 \lambda \vec{q} \, ^2}{3} (\alpha_2 + \beta_-^\prime) (p_0 + m_p) \nonumber \\
K_{1-} &= K_{1+} + \lambda ( - \alpha_0 + \beta_+ ) (p_0 - m_p) \nonumber \\
K_{2-} &= \frac{2 \lambda \vec{q} \, ^2}{3} (-\alpha_2 + \beta_+^\prime) (p_0 - m_p) \, \, \, ,
\end{align*}
where
\begin{align*}
\alpha_0 &= \frac{m_R^2 \alpha_1}{W^2} - 2\alpha_2 \vec{q} \, ^2 \, \, \, , \nonumber \\
\beta_+ &= (\frac{m_R^2 \beta_1}{W^2} - 2\beta_2 \vec{q} \, ^2) (W+m_p) \, \, \, , \nonumber \\
\beta_- &= (\frac{m_R^2 \beta_1}{W^2} - 2\beta_2 \vec{q} \, ^2) (W-m_p) \, \, \, , \nonumber \\
\beta_+^\prime &= \beta_2 (W+m_p) \, \, \, , \nonumber \\
\beta_-^\prime &= \beta_2 (W-m_p) \, \, \, ,
\end{align*}
and
\begin{equation*}
\lambda=\frac{g_{\pi N R}^2}{96 \pi W m_p^2 (m_R^2-W^2)} \, \, \, .
\end{equation*}
The contributions to all other partial waves vanish. The quantities $\alpha_1$, $\alpha_2$, $\beta_1$, and $\beta_2$ are defined by Eqs.~(\ref{eq:EQ031}) and (\ref{eq:EQ035}). As in the case of the Rarita-Schwinger propagator, 
the $(m_R^2-W^2)$ factor in the denominator of $\lambda$ introduces a pole only in $K_{1+}$, not in the other partial waves.

For the $s$-channel contributions, $K_{l\pm}^{3/2}=(\alpha + \beta) K_{l\pm}$ and $K_{l\pm}^{1/2}=(\alpha - 2 \beta) K_{l\pm}$.

The $u$-channel expressions comprise two contributions: those given by Eqs.~(42) of Ref.~\cite{mr}, with
\begin{equation*}
\lambda^\prime=\frac{g_{\pi N R}^2}{96 \pi W m_p^2}
\end{equation*}
and energy-dependent arguments ($a$,$b$)=($m_R^2 - m_p^2 - m_c^2 + 2 p_0 q_0$,$2\vec{q} \, ^2$) in all $\phi_n$ of Ref.~\cite{mr}, and the ones below, corresponding to the non-constancy of the coefficients $\alpha_1^{\prime \prime}$ 
and $\beta_1^{\prime \prime}$ in Eqs.~(\ref{eq:EQ092}) and (\ref{eq:EQ093}), respectively.
\begin{align*}
K_{0+} &= - \frac{\lambda^\prime}{2} (c_1 \phi_0 - c_2 \phi_1) \nonumber \\
K_{1+} &= - \frac{\lambda^\prime}{4} \big( 2 c_1 \phi_1 + c_2 (\phi_0 - 3\phi_2) \big) \nonumber \\
K_{1-} &= - \frac{\lambda^\prime}{2} (c_1 \phi_1 - c_2 \phi_0) \nonumber \\
K_{2+} &= \frac{\lambda^\prime}{4} \big( c_1 (\phi_0 - 3\phi_2) - c_2 (3\phi_1 - 5\phi_3) \big) \nonumber \\
K_{2-} &= \frac{\lambda^\prime}{4} \big( c_1 (\phi_0 - 3\phi_2) + 2 c_2 \phi_1 \big) \nonumber \\
K_{3+} &= \frac{\lambda^\prime}{16} \big( 4 c_1 (3\phi_1 - 5\phi_3) + c_2 (3\phi_0 - 30\phi_2 + 35\phi_4) \big) \nonumber \\
K_{3-} &= \frac{\lambda^\prime}{4} \big( c_1 (3\phi_1 - 5\phi_3) - c_2 (\phi_0 - 3\phi_2) \big) \, \, \, ,
\end{align*}
where
\begin{align*}
c_1 &= \big( \alpha_1 - \beta_1 (W - m_p) \big) (p_0 + m_p) \, \, \, , \nonumber \\
c_2 &= \big( \alpha_1 + \beta_1 (W + m_p) \big) (p_0 - m_p) \, \, \, .
\end{align*}

It must be emphasised that the values of $\phi_n$ in the above-shown $K_{l\pm}$ contributions are not the same as those entering Eqs.~(42) of Ref.~\cite{mr}. Therein, the energy-dependent arguments 
($a$,$b$)=($m_R^2 - m_p^2 - m_c^2 + 2 p_0 q_0$,$2\vec{q} \, ^2$) were implied in all $\phi_n$; on the contrary, ($a$,$b$)=($2 p_0 q_0 - m_p^2 -m_c^2$,$2\vec{q} \, ^2$) must be used in the expressions listed in this appendix.

Finally, $K_{l\pm}^{3/2}=(\alpha - \beta) K_{l\pm}$ and $K_{l\pm}^{1/2}=(\alpha + 2 \beta) K_{l\pm}$ for the $u$-channel contributions.

\subsection{\label{sec:NonPole}Non-pole contributions}

Also in this part, the energy-dependent arguments ($a$,$b$)=($2 p_0 q_0 - m_p^2 - m_c^2$,$2\vec{q} \, ^2$) are implied in all $\phi_n$.

From Eqs.~(\ref{eq:EQ094}) and (\ref{eq:EQ096}), one obtains the following non-pole isoscalar contributions to the partial waves for $l \leq 3$.
\begin{align*}
K_{0+} &= \frac{1}{2} (2 c_1 - c_2 \phi_0 - c_4 \phi_1) \nonumber \\
K_{1+} &= \frac{1}{4} \big( - 2 c_2 \phi_1 + c_4 (\phi_0 - 3\phi_2) \big) \nonumber \\
K_{1-} &= \frac{1}{2} (2 c_3 - c_2 \phi_1 - c_4 \phi_0) \nonumber \\
K_{2+} &= \frac{1}{4} \big( c_2 (\phi_0 - 3 \phi_2) + c_4 (3\phi_1 - 5\phi_3) \big) \nonumber \\
K_{2-} &= \frac{1}{4} \big( c_2 (\phi_0 - 3 \phi_2) - 2 c_4 \phi_1 \big) \nonumber \\
K_{3+} &= \frac{1}{16} \big( 4 c_2 (3 \phi_1 - 5 \phi_3) - c_4 (3 \phi_0 - 30 \phi_2 + 35 \phi_4) \big) \nonumber \\
K_{3-} &= \frac{1}{4} \big( c_2 (3 \phi_1 - 5 \phi_3) + c_4 (\phi_0 - 3 \phi_2) \big) \, \, \, ,
\end{align*}
where
\begin{align*}
c_1 &= - \frac{g_{\pi N R}^2}{192 \pi W m_p^2} \big( 4 m_R + 6 m_p + \frac{\lambda_1-\lambda_2 (W - m_p)}{W^2} \big) (p_0 + m_p) \, \, \, , \nonumber \\
c_2 &= - \frac{g_{\pi N R}^2}{192 \pi W m_p^2} \big( \lambda_1 + \lambda_2 (W - m_p) \big) (p_0 + m_p) \, \, \, , \nonumber \\
c_3 &= \frac{g_{\pi N R}^2}{192 \pi W m_p^2} \big( 4 m_R + 6 m_p + \frac{\lambda_1+\lambda_2 (W + m_p)}{W^2} \big) (p_0 - m_p) \, \, \, , \nonumber \\
c_4 &= \frac{g_{\pi N R}^2}{192 \pi W m_p^2} \big( \lambda_1 - \lambda_2 (W + m_p) \big) (p_0 - m_p) \, \, \, .
\end{align*}
The quantities $\lambda_1$ and $\lambda_2$ are defined in Subsection \ref{sec:WLTotal}. $K_{l\pm}^{3/2}=\alpha K_{l\pm}$ and $K_{l\pm}^{1/2}=\alpha K_{l\pm}$ for the non-pole isoscalar contributions.

From Eqs.~(\ref{eq:EQ095}) and (\ref{eq:EQ097}), one obtains the non-pole isovector contributions to the partial waves. The $K_{l\pm}$ expressions of the non-pole isoscalar part may be used, with redefined coefficients $c_n$, 
according to the following expressions.
\begin{align*}
c_1 &= - \frac{g_{\pi N R}^2}{192 \pi W m_p^2} \big( 2 (W - m_p) + \frac{\lambda_1-\lambda_2 (W - m_p)}{W^2} \big) (p_0 + m_p) \nonumber \\
c_2 &= \frac{g_{\pi N R}^2}{192 \pi W m_p^2} \big( \lambda_1 + \lambda_2 (W - m_p) \big) (p_0 + m_p) \nonumber \\
c_3 &= \frac{g_{\pi N R}^2}{192 \pi W m_p^2} \big( - 2 (W + m_p) + \frac{\lambda_1+\lambda_2 (W + m_p)}{W^2} \big) (p_0 - m_p) \nonumber \\
c_4 &= \frac{g_{\pi N R}^2}{192 \pi W m_p^2} \big( - \lambda_1 + \lambda_2 (W + m_p) \big) (p_0 - m_p) \nonumber
\end{align*}
$K_{l\pm}^{3/2}=\beta K_{l\pm}$ and $K_{l\pm}^{1/2}=- 2 \beta K_{l\pm}$ for the non-pole isovector contributions.

\clearpage
\newpage
\section{\label{App:AppC}Partial-wave decomposition of the scattering amplitude obtained in Pascalutsa's method (non-pole contributions)}

Regarding the partial-wave decomposition of the scattering amplitude, the details are given in Ref.~\cite{mr}.

From Eqs.~(\ref{eq:EQ098}) and (\ref{eq:EQ100}), one obtains the following non-pole isoscalar contributions to the partial waves for $l \leq 3$.
\begin{align*}
K_{0+} &= \big( c_1 + c_3 (W - m_p) \big) (p_0 + m_p) - \frac{1}{3} \big( c_2 - c_4 (W + m_p) \big) (p_0 - m_p) \nonumber \\
K_{1+} &= \frac{1}{3} \big( c_2 + c_4 (W - m_p) \big) (p_0 + m_p) \nonumber \\
K_{1-} &= K_{1+} - \big( c_1 - c_3 (W + m_p) \big) (p_0 - m_p) \nonumber \\
K_{2-} &= - \frac{1}{3} \big( c_2 - c_4 (W + m_p) \big) (p_0 - m_p) \nonumber
\end{align*}
The contributions to all other partial waves vanish. In these expressions,
\begin{align*}
c_1 &= - \frac{g_{\pi N R}^2}{192 \pi W m_p^2 m_R^2} \big( 4 m_R (2 m_R + m_p) p_{0R} + 4 m_p (m_R^2 - m_R m_p - m_p^2) - 2 (4 m_R + 3 m_p) \vec{q} \, ^2 \big) \, \, \, , \nonumber \\
c_2 &= - \frac{g_{\pi N R}^2}{96 \pi W m_p^2 m_R^2} (4 m_R + 3 m_p) \vec{q} \, ^2 \, \, \, , \nonumber \\
c_3 &= - \frac{g_{\pi N R}^2}{192 \pi W m_p^2 m_R^2} (W^2 - m_p^2 - m_c^2 + 2 p_0 q_0) \, \, \, , \nonumber \\
c_4 &= - \frac{g_{\pi N R}^2}{96 \pi W m_p^2 m_R^2} \vec{q} \, ^2 \, \, \, .
\end{align*}
$K_{l\pm}^{3/2}=\alpha K_{l\pm}$ and $K_{l\pm}^{1/2}=\alpha K_{l\pm}$ for the non-pole isoscalar contributions.

From Eqs.~(\ref{eq:EQ099}) and (\ref{eq:EQ101}), one obtains the non-pole isovector contributions to the partial waves. The $K_{l\pm}$ expressions of the non-pole isoscalar part may be used, with redefined coefficients $c_n$, 
according to the following expressions.
\begin{align*}
c_1 &= - \frac{g_{\pi N R}^2}{192 \pi W m_p^2 m_R^2} ( 2 m_R + 3 m_p ) (W^2 - m_p^2 - m_c^2 + 2 p_0 q_0) \nonumber \\
c_2 &= - \frac{g_{\pi N R}^2}{96 \pi W m_p^2 m_R^2} (2 m_R + 3 m_p) \vec{q} \, ^2 \nonumber \\
c_3 &= \frac{g_{\pi N R}^2}{192 \pi W m_p^2 m_R^2} ( 4 m_R m_p - 4 m_R p_{0R} + 12 m_p^2 + 10 \vec{q} \, ^2 ) \nonumber \\
c_4 &= - \frac{g_{\pi N R}^2}{96 \pi W m_p^2 m_R^2} 5 \vec{q} \, ^2 \nonumber
\end{align*}
$K_{l\pm}^{3/2}=\beta K_{l\pm}$ and $K_{l\pm}^{1/2}=- 2 \beta K_{l\pm}$ for the non-pole isovector contributions.

\end{document}